\documentclass[12pt]{iopart}
\usepackage{epsfig}
\usepackage{upgreek}

\begin{document}

\title[dos Santos Rolo et al.]{A Shack-Hartmann sensor for single-shot multi-contrast imaging with hard X-rays}

\author{Tomy dos Santos Rolo$^1$, Stefan Reich$^1$, Dmitry Karpov$^2$, Sergey Gasilov$^{3,4}$, Danays Kunka$^1$, Edwin Fohtung$^{2,6}$, Tilo Baumbach $^{1,7}$ and Anton Plech$^1$}

\address{$^1$Karlsruhe Institute of Technology (KIT), Institute for Photon Science and Synchrotron Radiation (IPS), Herrmann-von-Helmholtz-Platz 1, 76344 Eggenstein-Leopoldshafen, Germany.}
\address{$^2$New Mexico State University (NMSU), Department of Physics, 1255 N Horseshoe, Las Cruces, NM 88003-8001, USA.}
\address{$^3$Physical-Technical Institute, Tomsk Polytechnic University (TPU), Lenin Avenue 30, 634050 Tomsk, Russia.}
\address{$^4$Current affiliation: Canadian Light Source, 44 Innovation Boulevard Saskatoon, S7N 2V3, Canada.}
\address{$^5$Karlsruhe Institute of Technology (KIT), Institute of Microstructure Technology, Hermann-von-Helmholtz Platz 1, 76344 Eggenstein-Leopoldshafen, Germany.}
\address{$^6$Los Alamos National Laboratory, Los Alamos, NM 87545, USA.}
\address{$^7$Karlsruhe Institute of Technology (KIT), Laboratory for Applications of Synchrotron Radiation (LAS), Engesserstrasse 15, 76131 Karlsruhe, Germany.}

\ead{tomy.rolo@googlemail.com}
\ead{anton.plech@kit.edu}

\begin{abstract}

An array of compound refractive X-ray lenses (CRL) with 20x20 lenslets, a focal distance of 20 cm and a visibility of 0.93 is presented. It can be used as a Shack-Hartmann sensor for hard X-rays (SHARX) for wavefront sensing and  permits for true single-shot multi-contrast imaging the dynamics of materials with a spatial resolution in the micrometer range, sensitivity on nanosized structures and temporal resolution on the microsecond scale. The object's absorption and its induced wavefront shift can be assessed simultaneously together with information from diffraction channels. This enables the imaging of hierarchical materials. In contrast to the established Hartmann sensors the SHARX has an increased flux efficiency through focusing of the beam rather than blocking parts of it. We investigated the spatiotemporal behavior of a cavitation bubble induced by laser pulses. Furthermore, we validated the SHARX by measuring refraction angles of a single diamond CRL, where we obtained an angular resolution better than 4 $\upmu$rad.

\end{abstract}

\maketitle

\section{Introduction}

Assessing distinct features on different length scales in hierarchical materials typically involves the sequential application of different imaging techniques, impeding the simultaneous investigation of more than a single hierarchy level at once. Examining the dynamics of the evolution of hierarchical materials on the microsecond time scale has therefore been a considerable challenge. Furthermore, when the interrelation of dynamics on length scales from the millimeter down to the nanometer range is of interest, simultaneous hierarchical imaging  is mandatory.

The broad availability of X-ray sources such as synchrotrons and free electron lasers with very high brightness paved the way to tackle the formidable demands of imaging the structural evolution of hierarchical materials by means of multi-contrast imaging. 
Phase contrast imaging modalities assess the relative variations in the real part of the refractive index (propagation-based phase contrast \cite{Wilkins1996, Diemoz2012, Cloetens1996}, crystal-based \cite{Bonse1965} and grating-based interferometry \cite{Momose2003, Weitkamp2005, Pfeiffer2006, Momose2009, Pfeiffer2008}). 
Grating based interferometry and diffraction-enhanced imaging can additionally provide scattering contrast, thereby extending the accessible information to the full set of scalar wave-matter interactions in transmission geometry \cite{Kagias2017}.
However, the requirement to record several sub-images interrupted by mechanical motion (either with different sample positions, different relative positions of the gratings, or different angles of the analyzer crystal) complicate the investigation of dynamics on the microsecond scale \cite{McDonald2009}. Successful approaches have used a canted interferometer setup to retrieve the phase shifts by Fourier analysis of the Moir\'e pattern \cite{Momose2009} and could use polychromatic X-rays \cite{Momose2011}.

Hartmann sensors for X-rays provide single-shot differential phase-front measurements \cite{Wen2010, Morgan2011, Rutishauser2012}. By using absorption gratings, however, a large fraction of the incident radiation is absorbed in the optical element and can't contribute to the signal and therefore reducing the achievable signal-to-noise ratio for a given integration time.

We show the application of the analogue of a Shack-Hartmann sensor in the hard X-ray spectral range (SHARX), which permits for single-shot imaging of absorption, phase and diffraction contrast. This is quantitatively equivalent to the Hartmann-mask approach \cite{Wen2010, Morgan2011, Rutishauser2012, Rand2011, Vittoria2015SR}, or even conceptually similar to methods with structured X-ray beams \cite{Wang2015PRL, Wang2016, Kagias2016}. Due to the use of focusing lenslets \cite{Snigirev1996, Lengeler1998, Lyubomirskiy2016} the local X-ray flux density is increased strongly, leading to an improved signal-to-noise ratio even at shortest integration times. The availability of  diffraction contrast extends the accessible length scales \cite{Lynch2011, Strobl2014}. With a spatial resolution of the pitch of the SHARX (here 50 $\upmu$m) structures in the nanometer up to sub-micrometer regime can be imaged.

Since a single image is sufficient for obtaining the three contrast types simultaneously, contrast variations on the microsecond time scale can be used to judge how the structural dynamics on different length scales influence each other and the overall evolution of the material properties \cite{Ibrahimkutty2015}.


\section{Material and Methods}
\subsection{Fabrication of the 2D lens array}
\begin{figure}
	\includegraphics[width=\textwidth]{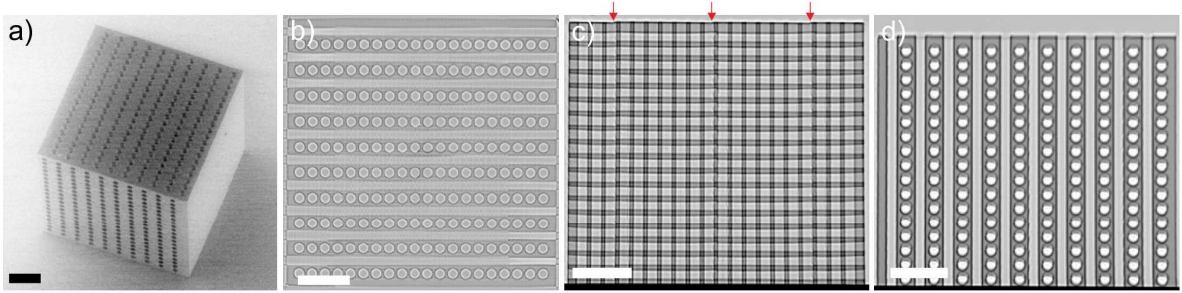}
	\caption{Images of the lens array: a) SEM image where the two planes of cylinder lenses can be seen (X-ray direction in use is perpendicular to the smooth surface), b) tomogram through a horizontal plane, c) and d) radiographs from the lateral and front side. In c) three vertical irregularities can be seen which originate from the printing process (marked with red arrows). Scale bars are 200 $\upmu$m.}
	\label{fig:SHARX}
\end{figure}

In Fig. \ref{fig:SHARX} we present the hard X-ray analogue to a visible-light array of lenslets in the shape of a 3D  hole structure fabricated from a low-absorbing polymer \cite{vonFreymann2010,Maruo1997}. The 3D structure with crossed hollow cylinders of 40 $\upmu$m  in diameter, a periodicity of 50 $\upmu$m  and a total volume of 1 mm$^3$ has been fabricated by using a state-of-the-art three-dimensional direct laser writing system (Photonic Professional GT, Nanoscribe GmbH). A commercial photoresist (IP-S, Nanoscribe GmbH) and a 25x magnification objective (0.8 NA) were used. The structure was fabricated in dip-in mode, meaning that the objective is immersed in the photoresist. After the writing process, the structure was developed in mr-Dev 600 (micro resist technology GmbH).

Each cylinder acts as a one-dimensional lens and the serial stacking shortens the focal length of the compound refractive lens (CRL)\cite{Snigirev1996, Lyubomirskiy2016}. The orthogonal combination of crossed cylinders in the beam propagation direction leads to a focusing of the beam in two directions. This results in well-defined spots that can be used analogously to Hartmann-mask beamlets. The benefits of this arrangement are manifold. The majority of photons from a lenslet aperture are concentrated on the beamlet area leading to a reduced exposure time due to the increased local photon flux density compared to the unfocused primary beam when using gratings or Hartmann-masks. Furthermore, the differential phase sensitivity can be adjusted to a certain extent by changing the sample-to-detector distance. The required lenslet pitches of tens of micrometers for hard X-ray diffraction experiments permit to vary cylinder diameters, grating materials and fabrication methods, and therefore tune the respective optical element for experimental demands. In particular, diffraction contrast is sensitive to length scales that can be tuned by the distance of the object to the detection plane and the array pitch \cite{Lynch2011,Strobl2014}.

\subsection{Experimental setup}
\label{sec:exp}
\begin{figure}
	\includegraphics[width=\textwidth]{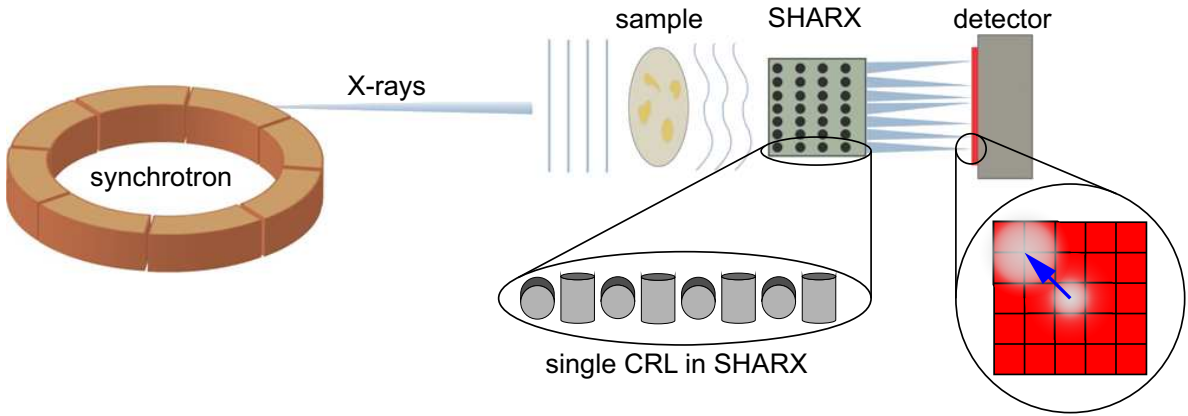}
	\caption{Schematic illustration of the beamline setup with the flat x-ray beam coming from the synchrotron getting disturbed by the sample and consequently analyzed by the SHARX and measured by the detector. The crossed cylinder lenses lead to small spots on the detector which may get shifted, broadened or/and attenuated as depicted by the blue arrow.}
	\label{fig:setup}
\end{figure}

The experiments were performed at the bending magnet beamline TOPO-TOMO at the synchrotron at KIT (Karlsruhe, Germany). The scheme of the Shack-Hartmann setup is shown in Fig. \ref{fig:setup}. A white beam (filtered by 0.2 mm of aluminum and 0.25 mm of beryllium), which has a beam size defined by mechanical blades (motorized slits) illuminates the lens array homogeneously, which leads to the formation of the beamlets on an X-ray scintillator. The visible light from the scintillator is imaged by a lens system onto a 2D detector. The sample can either be placed before or after the SHARX, which essentially determines the sensitivity (and susceptibility to phase wrapping) on phase contrast and on the length scale in diffraction contrast. Several setups for the imaging and optical characterisation were employed:

(i) The internal microstructure of the SHARX was examined by means of micro-tomography acquired in white beam with a 7 cm distance between the SHARX and the scintillator, which is much smaller than the focal distance of the lens array. A PCO.dimax camera was used to image a 10 $\upmu$m thick LSO:Tb scintillator by a 10x objective (Mitutoyo LWD 10X 0.28) and a 180 mm tube lens, resulting in a ninefold optical magnification. The effective pixel size was 1.22 $\upmu$m. Projections were continuously recorded while rotating the sample with constant angular velocity around an axis perpendicular to the beam propagation directions \cite{Rolo2014}.

(ii) The focal distance of the SHARX was characterized in monochromatic beam (double-multilayer monochromator with 2\% bandwidth at 9 keV). As detector an Andor Neo camera coupled to a 200 $\upmu$m LuAG:Ce scintillator via a single objective lens (Nikon Nikkor 85/1.4) was used. The optical magnification was 3.6x, resulting in an effective pixel size of 1.8 $\upmu$m.

(iii) The phase-front reconstruction of the diamond lens was performed in monochromatic beam (8.5 keV). The camera used was the Andor Neo using the same optics as in setup (i) resulting in an effective pixel size of 0.72 $\upmu$m. The distance between the detector and the SHARX was 16 cm and the diamond lens was placed before the SHARX.

(iv) The fast multi-contrast imaging during the ablation process was performed in white beam. The camera was a PCO.dimax, lens-coupled with twofold magnification to a 50 $\upmu$m  thick LuAg:Ce scintillator. The effective pixel size was 5.5 $\upmu$m. For each laser pulse an image sequence was recorded with a frame rate of 15 kHz and an exposure time of 33 $\upmu$s to limit motion blurring. The ablation process occurred in an in-situ chamber, which was placed behind the SHARX close  to the scintillator (7 cm distance). The ablation, in brief, was performed on a silver wire target (0.7 mm diameter) continuously transported through a sealed chamber of 0.4 cm$^3$ volume and flushed continuously by a water flow to avoid laser shielding by nanoparticles produced at predecessing laser pulses. Pulses from a nanosecond laser (Continuum Minilite I, 1064 nm, 10 mJ) entered the chamber via a lens (38 mm effective focal length in water) to be focused onto the target (for more details see \cite{Ibrahimkutty2015, Reich2017JCIS, Reich2017CPC}). For statistical reasons images from 700 subsequent laser shots were averaged.

\subsection{Image formation}

Every lenslet focuses a part of the incident plane wave onto a spot on the detector. After passing the object under investigation, each beamlet is attenuated, displaced or broadened relative to the incident beam direction (see Fig. \ref{fig:setup}) depending on the local transmission, refraction or diffraction properties of the sample, respectively. By using the central result from Fourier optics that a slice of sample placed in the path of a converging spherical wave produces the optical Fourier transform of the part of the object intersected by the wavefront \cite{Goodman2005}, it is possible to intuitively relate the local object properties to the modification of intensity in the focal plane. The angular spectrum of a planar wavefront is a delta function, thus the spots of the unmodified incident beamlets are as sharp as possible, where its size only depends on the optical transfer function of the lenslet aperture.
We have to note that the wavefront in reality is not necessarily planar due to the finite distance and lateral extent of the X-ray source. In this case the smallest achievable focal spot is the demagnified image of the source. Furthermore, the focal distance changes according to optical equations. This effect, however, is  small in our case as the distance from the source (30 m) is two orders of magnitude larger than the focal distance.

In the case of local attenuation in the sample, the direction of the wavefront is not changed, only the magnitude of the local intensity is reduced. Refraction, on the other hand, introduces a local change of direction of the planar wavefront without changing the frequency content of the angular spectrum. The apparent width of the resulting intensity distribution in the focal plane is unchanged compared to the unmodified incident beamlets. Any other wavefront curvature may arise from local scattering on sub-aperture-sized discontinuities of the refractive index in the sample \cite{Vittoria2015SR,Malecki2012}. These have a high-frequency component in the angular spectrum. Since the intensity response is the convolution of the optical transfer function with the angular spectrum of the incident wavefront \cite{Lynch2011,Vittoria2015SR}, this introduces a broadening of the focal spot. 

Small-angle scattering of objects that fulfil the far field condition adds to the broadening. Here, diffraction contrast averages signals from all the objects that fall into the sensitivity size range \cite{Lynch2011,Strobl2014}. By accessing several Fourier orders a crude size discrimination may be achieved \cite{Wen2009}. In practice, the distinction between phase contrast and diffraction contrast originating from scattering is not strict and depends on the geometry of the experimental setup, as well as discretization effects. In the continuous case, when the lenslet separation tends to zero, the distinction vanishes \cite{Koenig2016}.

In summary, the array of CRLs acts as the X-ray equivalent of a Shack-Hartmann sensor konwn in the optical regime \cite{Artzner1992}, which allows for discrimination of absorption contrast, phase contrast and additionally a set of diffraction contrasts depending on the geometry and detector resolution. Thus, a second level of size information on the nanometer scale is overlaid on the morphology on the micrometer scale.

\subsection{Retrieval of the contrasts: amplitude, phase and diffraction}
\label{sec:phaseretrival}

To retrieve the different contrasts of amplitude, differential phase and diffraction, each spot was fitted separately by a 2D-Gaussian:
\begin{equation}
  g = h \cdot exp \left\lbrace - \frac{1}{2} \cdot \left[\left(\frac{x- \mu_x}{\sigma_x^2}\right)^2 + \left(\frac{y-\mu_y}{\sigma_y^2}\right)^2\right] \right\rbrace + o ~,
\end{equation}
where h is the height, o the offset, $\sigma_{x,y}$ the width and $\mu_{x,y}$ the peak position in horizontal and vertical direction, respectively. Any rotation of the Gaussian distribution was neglected to reduce calculation time and because no pronounced elliptical spot deformation was detected. The offset was only calculated for better fitting, but represents no separate contrast as long as it is just the residual background intensity. 

To determine the changes in spot properties due to the sample the Gaussian fit was performed for two images, one without the sample and one with the sample. This leads to six contrast images for the undisturbed (free) image (height: $h_f$, offset: $o_f$, width in horizontal and vertical direction: $\sigma_{x,f}$ and $\sigma_{y,f}$, position in horizontal and vertical direction: $\mu_{x,f}$ and $\mu_{y,f}$) and equivalent for the sample image ($h_s$, $o_s$, $\sigma_{x,s}$, $\sigma_{y,s}$, $\mu_{x,s}$ and $\mu_{y,s}$). The changes where calculated by referring the undisturbed to the sample image for the transmission T, the differential phase dP and the diffraction D in horizontal and vertical direction by the following expressions:
\begin{eqnarray}
  T &=& \frac{h_s}{h_f}\\
  dP_x = \mu_{x,s} - \mu_{x,f} &and& dP_y = \mu_{y,s} - \mu_{y,f} \\
  D_x = \sigma_{x,s} - \sigma_{x,f} &and&  D_y = \sigma_{y,s} - \sigma_{y,f}
\end{eqnarray}
For calculating the actual beam deflection the differential phase was (in small-angle approximation) divided by the working distance. The latter being the distance between the SHARX or the detector when the sample is placed before the SHARX and the distance between the sample and the detector when the sample is placed between the SHARX and the detector.

Gaussian fits were used instead of Fourier-analysis, since this procedure prevents issues with phase wrapping \cite{Vittoria2015SR}. Additionally, the Fourier analysis as introduced by Wen et al. \cite{Wen2010} is known to introduce cross-talk between absorption and diffraction contrast \cite{Vittoria2015APL}. However, depending on the numbers of pixels sampling each beamlet, the Fourier analysis is able to probe several Fourier orders, possibly yielding information on several length scales in diffraction contrast as mentioned above.

For wavefront sensing the angular resolution is of primary importance. It depends of the spatial resolution of the detector and the working distance. In setup (i) of section \ref{sec:exp} a resolution of 1.8 $\upmu$m / 20 cm = 9 $\upmu$rad is expected for a shift of a full pixel. Nevertheless, peak shifts smaller than one pixel can easily be detected, leading to a  higher angular resolution.

The reconstruction of the change in the wavefront was done by zonal wavefront estimation introduced by Southwell \cite{Southwell1980}. For better accuracy in phase estimation the modified Southwell algorithm, as proposed in \cite{Pathak2014JO} was used. Here, additionally to the horizontal and the vertical phase slope, also the diagonal ones are incorporated in the reconstruction.

\section{Results and Discussion}

\subsection{Lens array characterization}
\begin{figure}
	\includegraphics[width=\textwidth]{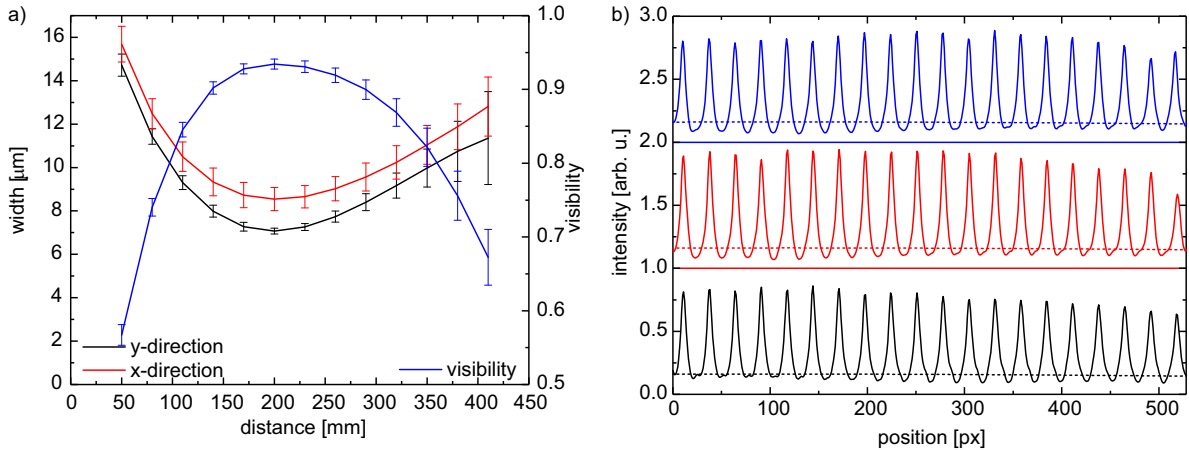}
	\caption{a) Spot sizes for the scanned distance between lens array and detector. The focal distance (average of all 400 spots) is about 200 mm (for 9 keV), defined where spots have the smallest width. The spots have a slightly different width for the x- and y-direction, which arises from different source sizes in horizontal and vertical axis. b) Line plots through three spot rows (red and blue offset shifted), showing the intensity profile in x-direction at a distance of 200 mm as well as the incoming X-ray intensity (dashed lines).}
	\label{fig:focus}
\end{figure}
	
	The focal length of biconcave X-ray lenses is calculated according to the basic optics formula \cite{Snigirev1996}:
\begin{equation}
	f = \frac{R}{2 \cdot N \cdot \delta} ~,
\end{equation}
where R is the radius of curvature, N the number of stacked biconcave lenses and 1-$\delta$ the energy-dependent real part of the index of refraction of the material in the X-ray range. Here, each CRL consists of 20 crossed cylindrical lenses which leads to an effective total of 10 biconcave lenses for each direction. The radius R is 20 $\upmu$m and $\delta = 5.65 \cdot10^{-6}$ at 9 keV. The expected focal length therefore is 177 mm. Scanning the distance between the detector and the array (see Fig. \ref{fig:focus} a)) yields the minimal width of the focal spots at (200 $\pm$ 15) mm which is in good agreement with the design value. Note that the real composition of the photoresist is unknown and therefore the exact value of $\delta$ may vary slightly. Additionally, lens errors may add to the shift. The slight difference in spot size in x- and y-direction originates from different sizes of the source in horizontal and vertical direction. The visibility V ($V = (I_{max}-I_{min})/(I_{max}+I_{min})$ with $I_{max}$ being the maximum and $I_{min}$ the minimum intensity in the square zone of each spot) of the spots peaks at the same distance as the effective focal length.

The achieved visibility of 0.93 is considerably higher than typical values for grating interferometry \cite{David2002}. Even in white-beam illumination (centered at 15 keV), we still obtained a high visibility of 0.5. Fig. \ref{fig:focus} b) shows three line plots in x-direction of different spot rows (solid lines, the upper two are shifted for clarity) in monochromatic mode. All of them show high gain factors compared to the free beam intensity (dashed lines). Two points have to be noted. First, not only intensity within the horizontal line is concentrated into the spots, but also in the perpendicular direction. Therefore intensity appears not to be conserved (area below dashed versus solid lines). Second, the minima in Fig. \ref{fig:focus} b) are not the global minima in each zone of the spot. The SHARX consists of crossed cylinders acting as 2D focusing lenses at the exact intersection point, while focusing works only in one direction at positions away from the cylinder cross points. The minima in Fig. \ref{fig:focus} b) reflect such positions leading to an increased background intensities compared to areas of no intensity increase or even intensity decrease. The true visibility, however, was derived from analysing a square area around each spot.  

\subsection{Reconstruction of the phase shift of a diamond lens}
\begin{figure}
	\includegraphics[width=\textwidth]{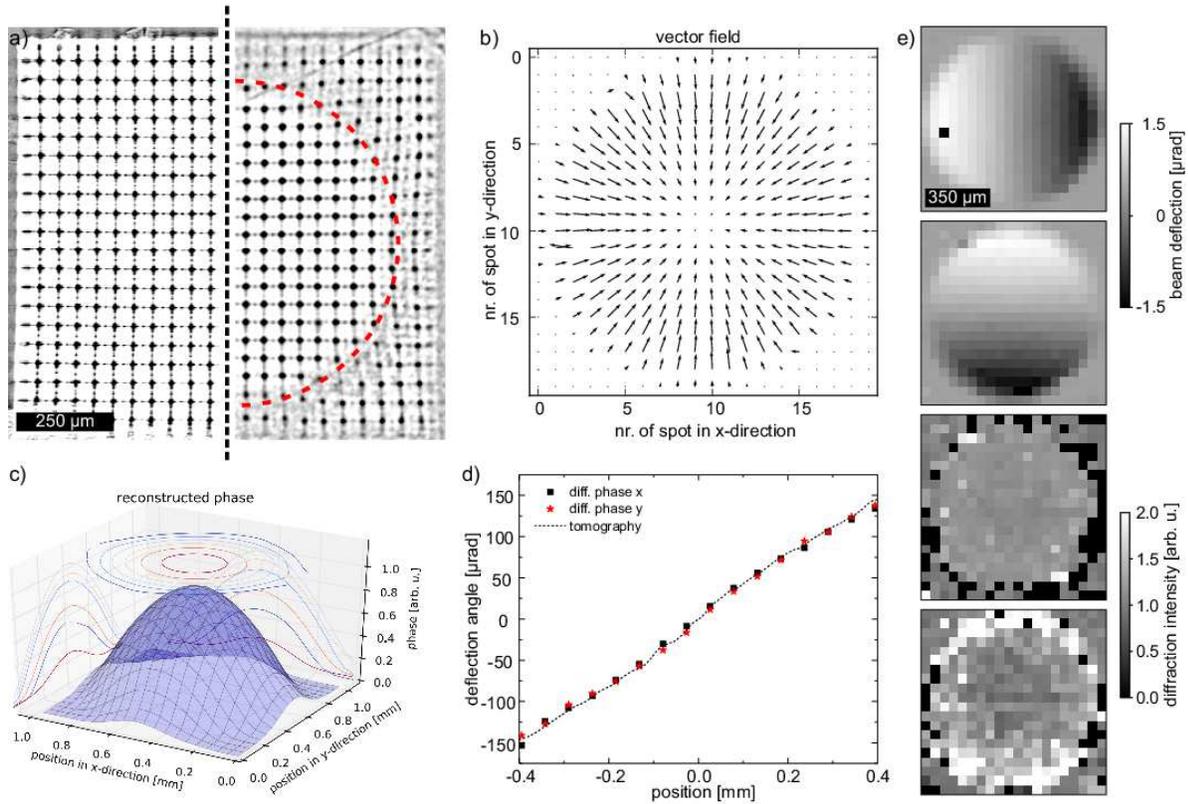}
	\caption{Reconstruction of the  phase shift of the beam introduced by a diamond lens. a) shows in the left half the unperturbed beamlet array and in the right half the pattern with diamond lens introduced (rim indicated by red dashed line, intensity inverted and contrasts optimized for better comparison). b) shows the vector field of the spot shifts. c) displays the reconstructed phase shift as surface plot with selected cuts in x, y and z. d) shows the deflection of X-rays along two orthogonal lines  crossing the CRL center (red and black markers) as derived by the Shack-Hartmann setup. A comparison with deflection angles computed based on the real lens shapes as determined by microtomography data (dashed line) is shown. e) shows (from top to bottom) the differential phase (as beam deflection) in x and y-direction and the diffraction in x and y-direction.}
	\label{fig:diamond}
\end{figure}

The SHARX can be used to analyze the changed wavefront of the X-ray beam in the exit plane of an X-ray optical element. For demonstration we analysed a CRL composed of 8 plano-concave parabolic lenses made from single-crystal diamond. Each unit lens has a design radius of curvature of 200 $\upmu$m at the vertex of the parabola and an entrance aperture of around 0.9 mm \cite{Polyakov2017,Gasilov2017OE} leading to a focal length of some 2.5 m (at 8.5 keV). Fig. \ref{fig:diamond} a) clearly illustrates the effect produced by the CRL inserted in the X-ray beam: the almost rectilinear grid of X-ray spots formed by the SHARX (left part) is bent toward the central axis of the CRL (right part). Shift in meridional and sagittal planes of each beamlet are derived by comparing reference and distorted image. The resulting local refraction angle in the CRL exit plane is shown as a vector field in Fig. \ref{fig:diamond} b). As expected, spots at the peripheral part of the CRL experience larger shifts than peaks located near the lens central axis. From these shifts we could easily reconstruct the X-ray phase shift using, for instance, the modified Southwell algorithm \cite{Pathak2014JO} (see also section \ref{sec:phaseretrival}). The calculated phase shift (Fig. \ref{fig:diamond} c)) is in reasonable agreement with the prediction based on the design values.

To verify the accuracy of the performed phase front metrology, the deflection of the X-rays introduced by the diamond CRL was derived independently from a microtomography (CT) with the CT rotation axis coinciding with the CRL optical axis\cite{Gasilov2017JSR}. The reconstructed distribution of material density was subjected to a segmentation procedure in order to outline the surfaces of the diamond. These in turn were used to calculate the thickness of diamond projected onto the CRL exit plane. Finally, this projected thickness was converted into X-ray deflection angles using the relation $\beta(x,y)=\Delta T(x,y)\cdot \delta_d$, where $\Delta T(x,y)$ is the finite difference approximation to the first derivative of the thickness and 1-$\delta_d$ the real part of the index of refraction of diamond. The result is shown as dashed line in Fig. \ref{fig:diamond} d) and compared to the SHARX deflection angles (red stars and black squares). Both angular shifts match well within a 4 $\upmu$rad confidence and relate to a CRL with effective radius of curvature at the apex of 205 $\upmu$m. A very good agreement between the measured beam deflection by the SHARX and the indirectly estimated deflection from the CT  demonstrates that a precise phasefront metrology can be performed with the SHARX device.

2D representations of the differential phase and diffraction contrast in x and y-direction are shown in Fig. \ref{fig:diamond} e). Phase contrast is shown for both directions, reproducing the finding of phase front curvature. The diffraction patterns show no significant signal as expected from a smooth material. However, the edges display an elevated diffraction level. Here two effects appear: the abrupt beam displacement within the individual beamlets, leading to a high curvature of the wave front, and an enhanced roughness due to processing. The first effect is in line with the interdependence of diffraction and phase from objects sized within the spatial resolution \cite{Yashiro2015}.

\subsection{In-situ imaging of laser-induced cavitation bubble}
\begin{figure}
	\includegraphics[width=\textwidth]{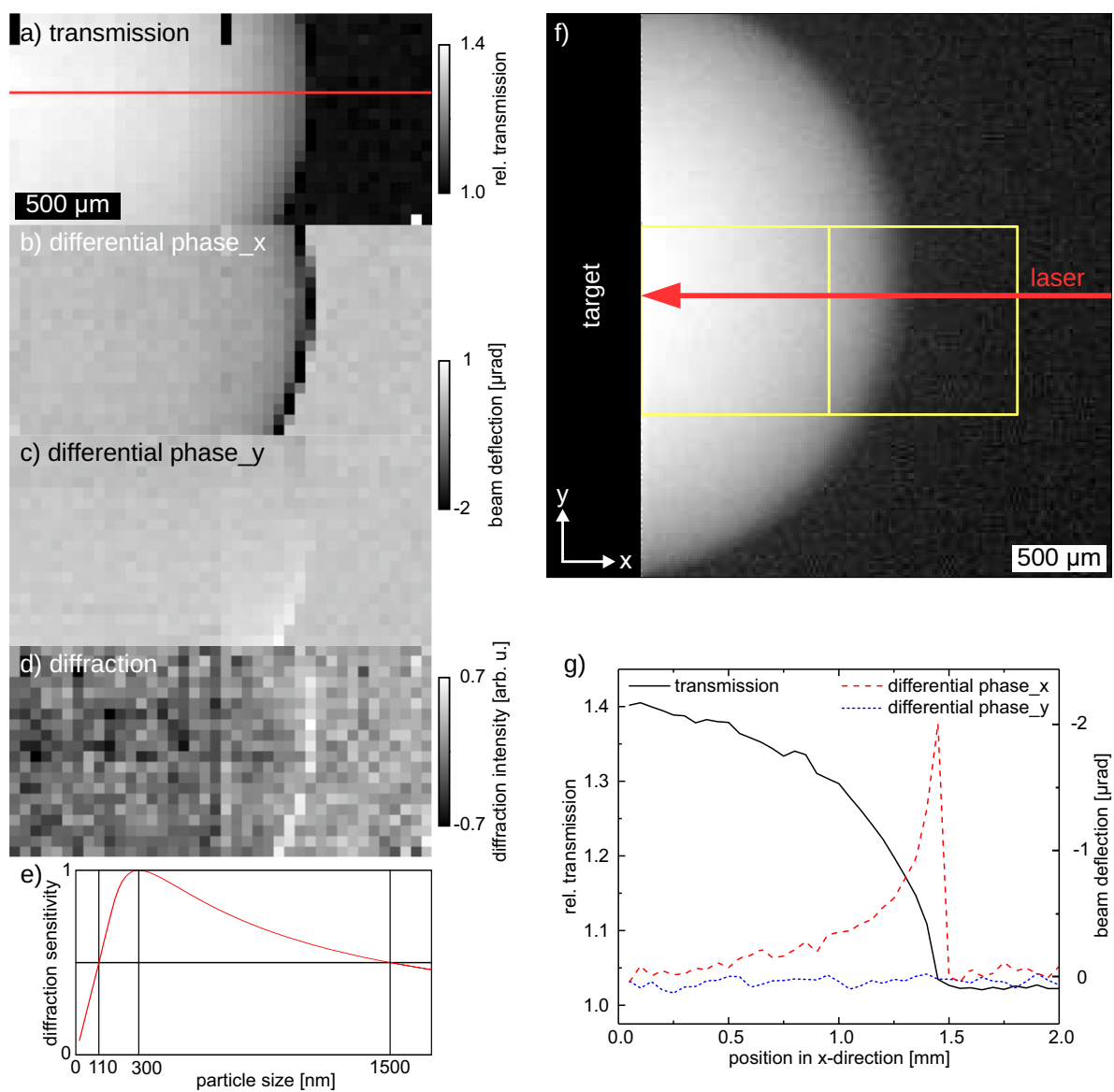}
	\caption{Results of a multi-contrast imaging of a transient cavitation bubble during ablation under liquid at its maximum expansion. Two separate measurements were repeated at two selected positions relative to the bubble, as indicated in the bright-field radiogram in f) and merged after reconstruction.
a) to d) show the different contrasts after averaging over 700 individual exposures of 33 $\upmu$s each. Note that the diffraction is the mean of the two directional values as no anisotropic scattering is expected. The graph in e) shows the calculated sensitivity curve on scattering particle size. The graph in g) shows clearly the correspondence of transmission and differential phase (shown in beam deflection) as function of bubble height above the target.}
	\label{fig:PLAL}
\end{figure}

Pulsed laser ablation in liquid (PLAL) is employed  to produce ligand-free nanoparticles in (e.g. aqueous) suspension \cite{Zeng2012}. Due to its hierarchical processes spanning several length scales and the fast dynamics of structure formation the  mechanisms and control of the process are still under investigation \cite{Zhang2017CR}. Within the present nanosecond excitation it is known that ablated material is ejected from the target by phase explosion, while at the same time a plasma is ignited to modify particle formation and foster energy coupling into the water. The latter is easily visible as a millimeter-sized transient cavitation bubble \cite{Reich2017CPC}. This bubble develops into a hemispherical shape of some 1.5 mm radius within a typical 150 $\upmu$s lifetime. It is known that this bubble contains a major part of the ejected mass from the target as nanoparticles \cite{Reich2017CPC,Reich2018APA} and the mutual interaction governs particle ripening \cite{Ibrahimkutty2012}. 

With the SHARX setup we could measure differential phase and diffraction in addition to the absorption and therefore gain information on the different hierarchical levels.
The bubble is imaged with the spatial resolution of the pitch of the SHARX (50 $\upmu$m). While the absorption and phase shift contain information about the bubble structure, the diffraction channel represents the scattering signal from structures in the nanometer up to the sub-micrometer scale as indicated by the sensitivity graph in fig. \ref{fig:PLAL}.
Due to the short exposure time an averaging over a number of (nominally reproducible) events is necessary for improving the signal-to-noise ratio.

Fig. \ref{fig:PLAL} f) shows an absorption contrast radiography without the SHARX for orientation \cite{Reich2017JCIS} over a large field of view. Due to the limited size of the SHARX array we sampled separately two different positions relative to the cavitation bubble, which were merged afterwards to cover the region of interest (see Fig. \ref{fig:PLAL} f)). Fig. \ref{fig:PLAL} a)-d) shows the results at maximum bubble expansion for the different contrast channels. The differential phases in both directions clearly show the rim of the bubble, which coincides with the highest phase gradient. The directional sensitivity is unambiguous. A reconstruction of the absolute phase could be calculated as for the diamond lens (not shown here).

As known from investigations with Hartmann masks, the diffraction channel is referred to the scattering of spatially unresolvable structures\cite{Wen2009, Wen2010, Morgan2011}. As explained in theory by Lynch et al.\cite{Lynch2011} the sensitivity for different particle sizes is mainly dependent on the pitch of the spots, the X-ray energy and the distance between the sample and the detector. The sensitivity peak for the present setup is located at a structure size of 300 nm with a FWHM from 110 to 1500 nm \cite{Lynch2011, Strobl2014} as shown in the sensitivity curve in Fig. \ref{fig:PLAL} e). From earlier studies it is known that the main size fraction of particles produced by PLAL is smaller than the present range\cite{Ibrahimkutty2015, Letzel2017}. However, ripening processes produce large particles that enter the visibility interval.

The small observed signals in the diffraction channel in fig \ref{fig:PLAL} d) (average of the two directional ones as no anisotropic scattering is expected) is of similar spatial distribution as the absorption signal. Therefore it is reasonable to consider a homogeneous filling of the bubble with scattering structures. On the other hand, some residual crosstalk from absorption or phase contrast \cite{Koenig2016, Kaeppler2014} may contribute to the signal and is subject of ongoing investigations.
The high diffraction signal at the bubble boundary is attributed to the crosstalk from an unresolvable sharp phase change \cite{Yashiro2015}.

{\textit A signal decomposition as in \cite{Kaeppler2014} could not shed more light on this issue. These signals are visible by Fourier analysis as well\cite{Vittoria2015SR}. A further setup optimization and data analysis are required.} It should be kept in mind that the integration time for the presented data only adds up to a total of 14 ms given the 33 $\upmu$s exposure of the individual frames.

\section{Summary and Conclusion}
Summarizing these results, the Shack-Hartmann approach with a 2D array of X-ray lenses, SHARX, is able to capture multi-contrast images of objects which discern absorption, phase and diffraction. This is quantitatively equivalent to the described Hartmann-mask approach \cite{Wen2010, Rand2011, Vittoria2015SR} or even conceptually similar to other methods with structured X-ray beams \cite{Wang2015PRL, Wang2016, Kagias2016}.  The only observable is the modification of a patterned X-ray field at the detector plane and its relation to the optical properties of the sample, but not to the way the patterning is produced. Sensitivity therefore relates to pure geometric design and pattern visibility.
The Shack-Hartmann sensor therefore increases usable X-ray flux compared to the Hartmann mask. An amplification factor of 6 has so far been achieved but with higher space filling fraction and at sources with smaller source size this can be improved further. Thus exposure times can be considerably lowered to study ultrafast processes. 
On the other hand, spatial resolution relative to the native detector pixel size is reduced. This can nevertheless be compensated for in some cases, where the sample resides closer to the detector (and is static). In that case the beamlets only illuminate a gridded part of the sample. A lateral rastering of the sample can then regain spatial resolution by interleaving several exposures. 

As a refraction-based element the SHARX is chromatic due to the fact that the refractive index of any lens material is energy-dependent. Consequently, the focal point will shift with X-ray energy. Nevertheless, this shift (estimated as 5.9 cm/keV at 9 keV) in our case is comparable to broadening due to the imaging of the source onto the detector as compared to the filtered spectrum with 50\% bandwidth. With quasi-monochromatic sources (some 5\% bandwidth) the broadening due to shifting the focal length would be about 0.2 $\upmu$m, which is small as compared to the source image in our case (7 $\upmu$m).

In this paper we have demonstrated the use of a 2D array of cylindrical holes inside a refractive polymer, called SHARX, as a compound refractive lens array for hard X-rays. We have produced a lens array of 50 $\upmu$m pitch with a focal length of 200 mm, which is sensitive to angular shifts of the wavefront down to 4 $\upmu$rad at a micrometer spatial resolution. At the same time structural length scales of some 300 nm are probed in diffraction contrast. The device requires low precision of alignment and no moving parts during the data acquisition. The performance is preserved at a relaxed bandwidth of the X-ray beam.

The sensitivity of phase modulation by an object (sample) has been shown to be very high and comparable to other phase-sensitive techniques such as the analyser-based wavefront sensing shown here. Adding the high flux, this approach is particularly useful for imaging irreversible or fast processes in materials or sensing wavefront (changes) of X-ray beams. The efficiency of the method can be further improved by increasing the filling factor for the lenses and focusing precision by shifting to individual parabolic lenses rather than lens cylinders.   

The availability of a diffraction contrast is able to fulfil demands imposed by the requirements of imaging hierarchical structure evolution. The simultaneously observable length scales combine the micrometer range in absorption contrast and the nanometer to sub-micrometer range in diffraction contrast at the spatial resolution of the absorption contrast. This results in the ability to assess contrast variations of interest on the microsecond time scale and to investigate how the structural dynamic influences each other.

\section*{Acknowledgments}
This research has been supported by the Helmholtz programme topic "from matter to materials and life - MML" and by the German Research Foundation DFG through grant PL 325/8-1. 
The Karlsruhe Nano Micro Facility KNMF is acknowledged for producing the lens array. We are grateful to TISNCM (Troitsk, Russian Federation) for providing  a single crystal diamond CRL for testing. The provision of beamtime at ANKA is acknowledged. We wish to thank T. M\"uller for support at the beamline and S. Barcikowski and A. Letzel for continuing discussions.

\end{document}